# Funding CRISPR: Understanding the role of government and philanthropic institutions in supporting academic research within the CRISPR innovation system

David Fajardo-Ortiz[1,2], Stefan Hornbostel[3], Maywa Montenegro de Wit[4], and Annie Shattuck[5]

[1]Research System and Science Dynamics Research Area, Deutsche Zentrum für Hochschul-und Wissenschaftsforschung (DZHW), Berlin, Germany
[2]Department of Cardiovascular Sciences, KU Leuven, Leuven, Belgium
[3]Institut für Sozialwissenschaften, Humboldt-Universität zu Berlin, Berlin, Germany
[4]Environmental Studies Department, University of California Santa Cruz, Santa Cruz, CA, USA
[5]Department of Geography, Indiana University, Bloomington, IN, USA

an open access journal

Citation: Fajardo-Ortiz, D., Hornbostel, S., Montenegro de Wit, M., & Shattuck, A. (2022). Funding CRISPR: Understanding the role of government and philanthropic institutions in supporting academic research within the CRISPR innovation system. *Quantitative Science Studies*, *3*(2), 443–456. https://doi.org/10.1162/qss_a_00187

DOI: https://doi.org/10.1162/qss_a_00187

Peer Review: https://publons.com/publon/10.1162/qss_a_00187

Received: 31 May 2021
Accepted: 1 March 2022

Corresponding Author:
David Fajardo-Ortiz
davguifaj@gmail.com

Handling Editor:
Ludo Waltman

## ABSTRACT

CRISPR/Cas has the potential to revolutionize medicine, agriculture, and biology. Understanding the trajectory of CRISPR research, how it is influenced, and who pays for it is an essential research policy question. We use a combination of methods to map, via quantitative content analysis of CRISPR papers, the research funding profile of major government agencies and philanthropic organizations and the networks involved in supporting key stages of high-influence research, namely, basic biological research and technological development. The results of the content analysis show how the research supported by the main U.S. government agencies focuses both on the study of CRISPR as a biological phenomenon and on its technological development and use as a biomedical research tool. U.S. philanthropic organizations, with the exception of HHMI, tend, by contrast, to specialize in funding CRISPR as a genome editing technology. We present a model of cofunding networks at the two most prominent institutions for CRISPR/Cas research (the University of California system and the Broad/Harvard/MIT system) to illuminate how philanthropic organizations have articulated with government agencies to cofinance the discovery and development of CRISPR/Cas. Our results raise fundamental questions about the role of the state and the influence of philanthropy over the trajectory of transformative technologies.

## 1. INTRODUCTION

CRISPR/Cas is a set of versatile technologies aimed to manipulate, analyze, and visualize the biomolecular machinery of living organisms (Pickar-Oliver & Gersbach, 2019). CRISPR/Cas has the potential to revolutionize medicine (Butler & Tector, 2017; Kannan & Ventura, 2015; Schermer & Benzing, 2019), agriculture (Gao, 2018), and the way we understand life itself (Ledford, 2015). The impact of CRISPR/Cas genome editing technologies has been recognized with the 2020 Nobel Prize in Chemistry awarded to Emmanuelle Charpentier and Jennifer A. Doudna "for the development of a method for genome editing" (Uyhazi & Bennett, 2021). Applications for these technologies have been proposed in fields as diverse as pharmaceuticals (Lu, Livi et al., 2017), crop development (Chen, Wang et al., 2019), livestock breeding (Petersen, 2017), industrial biotechnology (Donohoue, Barrangou, & May, 2018), and pest control (McFarlane, Whitelaw, & Lillico, 2018). As CRISPR/Cas represents one of the most potentially transformative technological breakthroughs of the last decade, it is important for researchers, policymakers, and the public to understand innovation trajectories, who finances them, who bears the risks and rewards of innovation, and for whom technologies are ultimately developed. CRISPR/Cas has been analyzed using myriad historical, legal, ethical, policy, and scientometric approaches. The social science and humanities discussion on CRISPR/Cas technologies ranges from ethical concerns about heritable genome editing (Reyes & Lanner, 2017) to intellectual property (Parthasarathy, 2018; Sherkow, 2017) and democratization and governance of these technologies (Montenegro de Wit, 2020). Despite the diversity of such studies, the role of financing on CRISPR/Cas research—specifically how different types of funding influence the research process—remains largely opaque.

Since the Second World War, in the period known as the golden age of capitalism, government agencies have been the main source of funding for scientific research in the U.S. academy (Geiger, 2008). Breakthrough technologies have emerged from long-term investments in R&D under the banner of a "public good" mission (Mazzucato, 2015). Amid the growing influence of productivist coalitions since the 1930s—farm commodity groups, land-grant administrators, agribusiness firms, and federal agricultural agencies (Buttel, 2005)—government agencies have largely taken the lead in actively shaping markets and shouldering the risk of early-stage transformative research investments (Mazzucato, 2015). Private-sector actors typically limited their role to lower risk forms of technology integration, development, and marketing later in the innovation process. This status, during a period of corporate consolidation, is less contradictory than it first appears. A schism between "basic" and "applied" science has long been recognized as favorable to private-sector interests who actively worked to carve a social division of labor that put universities in charge of basic research and positioned industry to control the commodity form (Kloppenburg, 2005). Though recognized as a taxpayer subsidy to agribusiness in the agricultural sector (Glenna, Lacy et al., 2007), this relationship has not only withstood time but has deepened: With the passage of Bayh-Dole in 1980, Congress fundamentally shifted the incentive structure governing research and development by allowing publicly funded institutions to own inventions resulting from federally sponsored research and to license those inventions to the private sector (Boettiger & Bennett, 2006).

Early research progress on genome technologies (viral vectors, RNAi, and the different genome editing platforms) largely followed this pattern, with the U.S. National Institutes of Health (NIH) playing a leading role in funding innovation over the past 30 years (Fajardo-Ortiz, Shattuck, & Hornbostel, 2020). Scientometrics studies on the impact of U.S. government institutions, such as the National Science Foundation and the NIH, show that these institutions still function as global driving forces of innovation writ large (Chen, Roco et al., 2013; Li, Azoulay, & Sampat, 2017).

However, there has been a clear decline in U.S. government support for science in recent years when measured as a percentage of gross domestic product (Boroush, 2020), while simultaneously a second golden age in the economic power of philanthropic and charitable organizations is taking place in the United States (Stevens, 2019). The emerging active role of philanthropic foundations as patrons of science has serious implications for the governance of science and technology. For example, Anne-Emanuelle Birn has documented the capacity of philanthropy to change the global research agenda on health from a focus on the social determinants of health to sophisticated technological solutions, with mixed results (Birn, 2014). Similarly, the participation of philanthropic organizations in the area of agriculture and food sciences is currently promoting the development and implementation of "silver bullet" technologies that allow the incorporation of farmers into commercial value chains (Brooks, 2013). Moreover, an investigation into the role of science philanthropy in U.S. research universities showed how this type of organization concentrates universities' efforts on the translation of knowledge from basic research to the development of applications, much like other private sector actors (Murray, 2013). Understanding how different types of funding accrue to and support different levels of research—from basic science to the development of technologies to their ultimate application in specific industries—can provide us with fundamental information on the influence of these different organizational actors on the development and implementation of CRISPR/Cas technologies.

The University of California system and the Broad/Harvard/MIT system[1] are the two most prominent academic institutions involved in the research and development of CRISPR/Cas technologies. The impact of these two research systems on the invention and development of CRISPR/Cas technologies has been well documented (Cockbain & Sterckx, 2020; Fajardo-Ortiz et al., 2020; Parthasarathy, 2018). Therefore, these institutions are an excellent case study to examine the evolution of these technologies, the funding networks that support them, and the relationships between innovation, financing, production, and property rights over these technologies. In the present investigation we aim to map and critically analyze the organization of the cofunding networks of highly cited research on CRISPR/Cas technologies performed by the UC and the Broad/Harvard/MIT systems. Importantly, while the University of California is a public institution, the Broad/Harvard/MIT system is made up of private institutions. This distinction is relevant for the study of university research funding, as differences have been reported in terms of the ability to access government and philanthropic funding between public and private universities (McPherson, Gobstein, & Shulenburger, 2010; Taylor, 2016; Zhang, 2019). Doing so raises fundamental questions about the articulation between government agencies and philanthropic organizations in supporting breakthrough innovations, risk, reward, and democratic influence over the trajectory of transformative technologies.

Our analysis focuses on the initial stages of the CRISPR/Cas innovation process that take place in academic institutions. In this sense, we focus on two types of research published in scientific journals (Table 2; see Section 2). The first type of research relates to the study of CRISPR as a biological phenomenon, that is, as a component of the bacterial immune system. This category is important because it is the source of the necessary scientific knowledge that allows the continuous development of CRISPR/Cas technologies. For example, the original invention of the CRISPR/Cas9 system as a genome editing technology (Jinek, Chylinski et al., 2012) was strongly based on a number of previous scientific discoveries, such as the role of small RNAs for sequence-specific detection and silencing of viral nucleic acids (Wiedenheft, Sternberg, & Doudna, 2012) or the role of RNase III in bacterial immunity (Deltcheva, Chylinski et al., 2011). Basic research on CRISPR/Cas as a biological phenomenon has not ended and continues to provide scientific knowledge for the subsequent development of CRISPR/Cas technologies. The second type of research is related to the continuous development of CRISPR/Cas technologies, that is, the set of investigations subsequent to the original invention of CRISPR/Cas9 genome editing system. The continuous development of CRISPR/Cas technologies is oriented to extending the use of CRISPR/Cas to different organisms and cell types, increasing the specificity of CRISPR/Cas, and the search for platform alternatives to Cas9 that could overcome the foundational intellectual property restrictions on prior inventions. These categories do not include research where CRISPR/Cas is used primarily to investigate or change the function of genes in target organisms, including analyzing the role of specific genes in normal or pathological biological processes and applications of CRISPR/Cas in biomedicine, food systems, environmental systems, or industrial biotechnology. For the purposes of this analysis, we also excluded papers reporting the ethical or societal implications of CRISPR/Cas.

[1] The Broad/Harvard/MIT system comprises Harvard University, the Massachusetts Institute of Technology, and the Broad Institute, an independent research nonprofit that partners with MIT and Harvard. For this study, we included research conducted at all three institutions in this system.

While basic research on CRISPR as a biological phenomenon provides the scientific knowledge necessary for the invention of CRISPR/Cas technologies, research related to the development of CRISPR/Cas technologies is central to downstream applications and IP rights, and thus beneficiaries of this technological revolution. In this regard, we seek to illuminate how philanthropic and charitable organizations have engaged with U.S. government agencies to cofinance the discovery and development of CRISPR/Cas technologies from a macroscopic perspective of major U.S. philanthropic and governmental funding sources, and from a perspective focused on research funding at the University of California and the Broad/Harvard/MIT systems. To get a macroscopic perspective, we coded the content of CRISPR publications funded by the four largest U.S. government agencies and four largest U.S. philanthropic organizations. For the focused perspective, we mapped the cofunding networks for highly influential research that took place at the University of California and the Broad/Harvard/MIT systems. This method may be useful as a research policy tool to allow decision makers to visualize influence on public sector agencies over the direction of technological development.

## 2. METHODOLOGY

To get a codified overview of the CRISPR research portfolios of major U.S. government agencies and philanthropies, we combined VOSviewer's scientometric analysis (Van Eck & Waltman, 2017) and KH Coder's text mining tools (Higuchi, 2016) as follows:

1. In February 2022, we conducted a search for papers on CRISPR in the Web of Science (WoS) (Mongeon & Paul-Hus, 2016) using the following criteria: CRISPR or "clustered regularly interspaced short palindromic repeats" (Topic) and Articles (Document Types). We found 22,834 papers.
2. With the bibliometric information from WoS, we built a co-occurrence network model of relevant terms related to CRISPR research (Figure 1). The clustering analysis identified six clusters in the co-occurrence network. The three largest clusters are related to biological and pathological processes mainly at the cellular level (i.e., CRISPR as a research tool (red nodes)); CRISPR as a component of the innate immune system of certain bacteria and archaebacteria (i.e., CRISPR as a biological phenomenon (blue nodes)); and CRISPR as a genome editing technology (green nodes). We coded the terms with the highest total link strength from the three main clusters into the following categories: phenomenon, technology, and research tool.
3. From WoS, the bibliometric information (title, abstract and keywords) of the research articles that report funding from the following entities was downloaded: National Institutes of Health (NIH): 6,218 papers; National Science Foundation (NSF): 1,092 papers; United States Department of Defense (DoD): 341 papers; U.S. Department of Energy (DoE): 255; Howard Hughes Medical Institute (HHMI): 329; Burroughs Wellcome Fund (BWF): 155 papers; Bill & Melinda Gates Foundation (BMGF): 111 papers; and the Welch Foundation (TWF): 104 papers.
4. The content (title, abstract and keywords) of the articles financed by these organizations was analyzed with KH Coder using the coding built in the previous steps. A heat map and clustering were generated by crossing the categories "phenomenon," "technology," and "research tool" with the sources of financing analyzed.

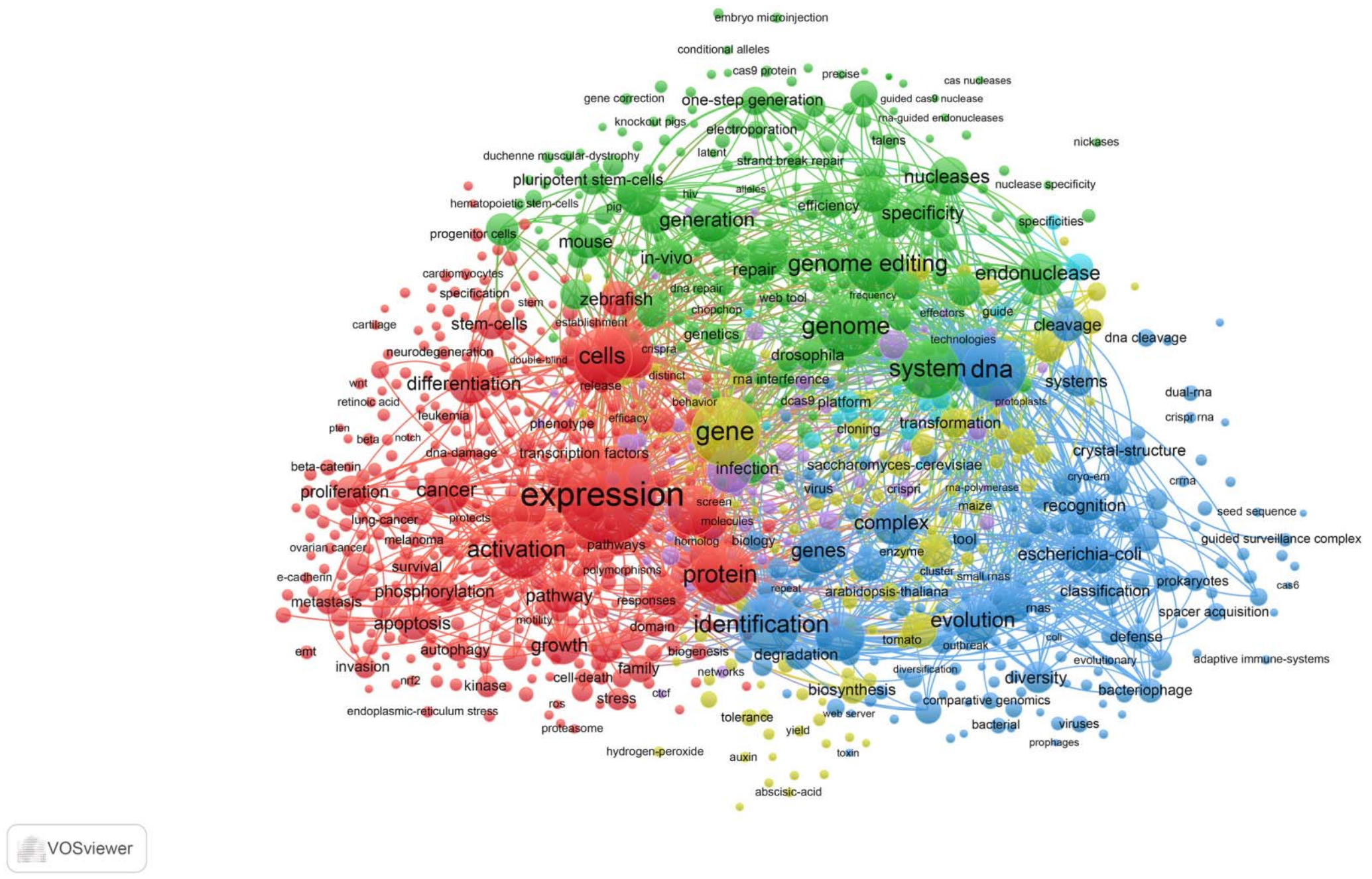

**Figure 1.** Co-occurrence map of relevant terms in articles on CRISPR.

**Table 1.** Search criteria, number of papers found in the Web of Science (WoS), number of papers selected and number of papers forming the cofunding network models. It is important to note that because of the historical relevance of Cas9 and Cas12a as the first two CRISPR platforms for genome editing (Ishino, Krupovic, & Forterre, 2018) we use them as additional terms in the search criteria, and the search for terms in WoS was carried out as a full-text search, not limited to terms in the title. Thus, although we did not explicitly search for the different nucleases (and their pseudonyms), any paper with a term associated with CRISPR will appear in the search. This scope is important because the two university systems have distinct IP rights over the Cas variants and we aimed to be as inclusive as possible.

| | |
|---|---|
| **Broad/Harvard/MIT System** | |
| Search criteria: | |
| **ORGANIZATION-ENHANCED**: (Broad Institute or Massachusetts Institute of Technology (MIT) or Harvard University) AND **TOPIC**: (CRISPR or "clustered regularly interspaced short palindromic repeats" or Cas9 or Cpf1 or Cas12a). Refined by: **DOCUMENT TYPES**: (ARTICLE) | |
| Total of papers found in the WoS (by April 25, 2020) | 922 |
| Total of papers selected from WoS with at least 25 citations (receiving 95.2% of citations) | 364 |
| **University of California System** | |
| Search criteria: | |
| **ORGANIZATION-ENHANCED**: (University of California System) AND **TOPIC**: (CRISPR or "clustered regularly interspaced short palindromic repeats" or Cas9 or Cpf1 or Cas12a). Refined by: **DOCUMENT TYPES**: (ARTICLE) | |
| Total of papers found in the WoS (by April 25, 2020) | 920 |
| Total of papers selected from WoS with at least 15 citations (receiving 94.8% of citations) | 400 |

**Table 2.** Research levels of the papers in the network models

| | |
|---|---|
| A. Biological phenomenon | Papers investigating CRISPR/Cas as bacterial immune systems; the interactions of their component in bacteria or archaea, and/or the molecular, ecological, or evolutionary interactions of these systems with the bacteriophage viruses. |
| B. Developments or improvements of CRISPR/Cas technologies | These papers report discoveries of alternative CRISPR/Cas genome editing systems that could be more efficient, more versatile, or easier to use; investigations aimed to overcome the technical difficulties to apply the technology in different organisms; or investigations reporting molecular mechanisms that can be used to modulate the activity of Cas enzymes. |
| C. Other | Papers in which CRISPR/Cas is not the central object of investigation but is an instrument to identify or analyze the role of specific genes in normal or pathological biological processes; papers reporting applications of CRISPR/Cas in biomedicine, food systems, environmental systems, or industrial biotechnology; or papers reporting the ethical or societal implications of CRISPR/Cas technologies. |

To build the funding network for highly cited research on CRISPR produced by the Universities of California and the Broad/MIT/Harvard systems, we follow these steps:

5. A search of peer-reviewed papers was performed in WoS in May 2020. The research criteria are listed in Table 1. A very similar number of papers was found for the UC system and the Broad/Harvard/MIT system (920 and 922 respectively; see Table 1). For each system we selected a number of top-cited papers that concentrated approximately 95% of the citations received. That is, we wanted to focus our explorative analysis on the most influential papers from each institutional system. Roughly 40% of top-cited papers accumulated 95% of citations, while the remaining 60% of papers received just 5% of citations (Table 1).
6. A bimodal network model of papers and cofunding organizations was built for each institutional system by using the information reported in the acknowledgment section of the papers.
7. The papers in the bimodal network models were classified into three different types of research (Table 2).
8. The cofunding organizations in the bimodal network models were classified as follows. A: U.S. government agencies—any public source of funding in the United States including federal, state, and local agencies, and armed forces. B: Philanthropic or charitable organizations (i.e., nonprofit organizations that are tax exempt under 501(c)(3) requirements in the United States (Bertrand, Bombardini et al., 2020)). C: Other organizations, such as academic institutions, professional organizations, medical research centers, and non-U.S.-based organizations.[2]
9. A bimodal network model of papers and cofunding organizations was built for each institutional system by using the information reported in the acknowledgment section of the papers. The cofunding network models were visualized and analyzed with Cytoscape

[2] Here it is necessary to clarify that Howard Hughes Medical Center, an influential node in our network model, is a center for basic and applied academic medical research and is often thought of primarily as a research institution. However, its most recent 990 (publicly available tax form for the period ending August 2019) shows $1.3 billion in investment income from its endowment, dwarfing the grants it received ($2.8 million) and its program service revenue ($2.6 million). It gave out $34 million in grants (ProPublica, 2019). Given its 501(c)(3) status and its financial statements, we included HHMI in the philanthropic and charitable organization category.

(Saito, Smoot et al., 2012). Only those papers that were classified as (A) research on CRISPR as a biological phenomenon and (B) research aimed at the development and improvement of CRISPR/Cas technologies were included in the bimodal network models. Likewise, only those institutions based in the United States classified as (A) government agencies and (B) philanthropic or charitable organizations were included in the network model. Finally, only those organizations that funded at least three CRISPR investigations (type A or B) from the previously selected sets of highly cited articles were included.

## 3. RESULTS

KH Coder generated a heat map and clustering from the tabulation of the terms coded in the categories "phenomenon," "technology," and "research tool" versus the different sources of financing analyzed (Figure 2). The heat map and clustering clearly identify three different profiles of entities that fund CRISPR research. A first profile is characterized by a relatively high percentage of articles with terms related to the categories "phenomenon" and "technology" (Figure 2). Such is the case for the DoE and the NSF, which appear aggregated in the same cluster. A second profile is characterized by a high percentage of articles with terms coded in the technology category and relatively lower percentages of the other two categories (Figure 2). The philanthropic organizations TWF, BMGF, and BWF follow this pattern. Finally, a third funding entity profile shows a balance between the terms related to the research tool and technology categories (Figure 2). Such is the case with the HIH, the DoD, and the HHMI.

A cofunding network model was built for each institutional system showing how philanthropic and charitable organizations articulated with U.S. government agencies to finance the discovery and development of CRISPR/Cas technologies that took place at the University of California and Broad/Harvard/MIT systems. The bimodal network model of the Broad/Harvard/MIT system was formed by 28 organizations (12 governmental agencies and 16 philanthropic/charitable organizations) and 111 highly cited papers (14 papers on CRISPR as a biological phenomenon and 97 papers on the development of CRISPR/Cas technologies; Figure 3). Almost half of this set of UC papers were authored by Jennifer Doudna (31 papers on CRISPR as a biological phenomenon and 26 papers reporting developments of CRISPR/Cas technologies). On the other hand, the bimodal network model of the UC system was formed by 15 funding organizations (10 governmental agencies and five philanthropic/charitable organizations) and 117 highly cited papers (59 papers on CRISPR as a biological phenomenon and 58 papers on the development of CRISPR/Cas technologies; Figure 3). Feng Zhang is the author of a quarter (28 of 111) of the papers on the bimodal cofinancing network model of the Broad/Harvard/MIT system while David R. Liu and George Church each authored 14 papers.

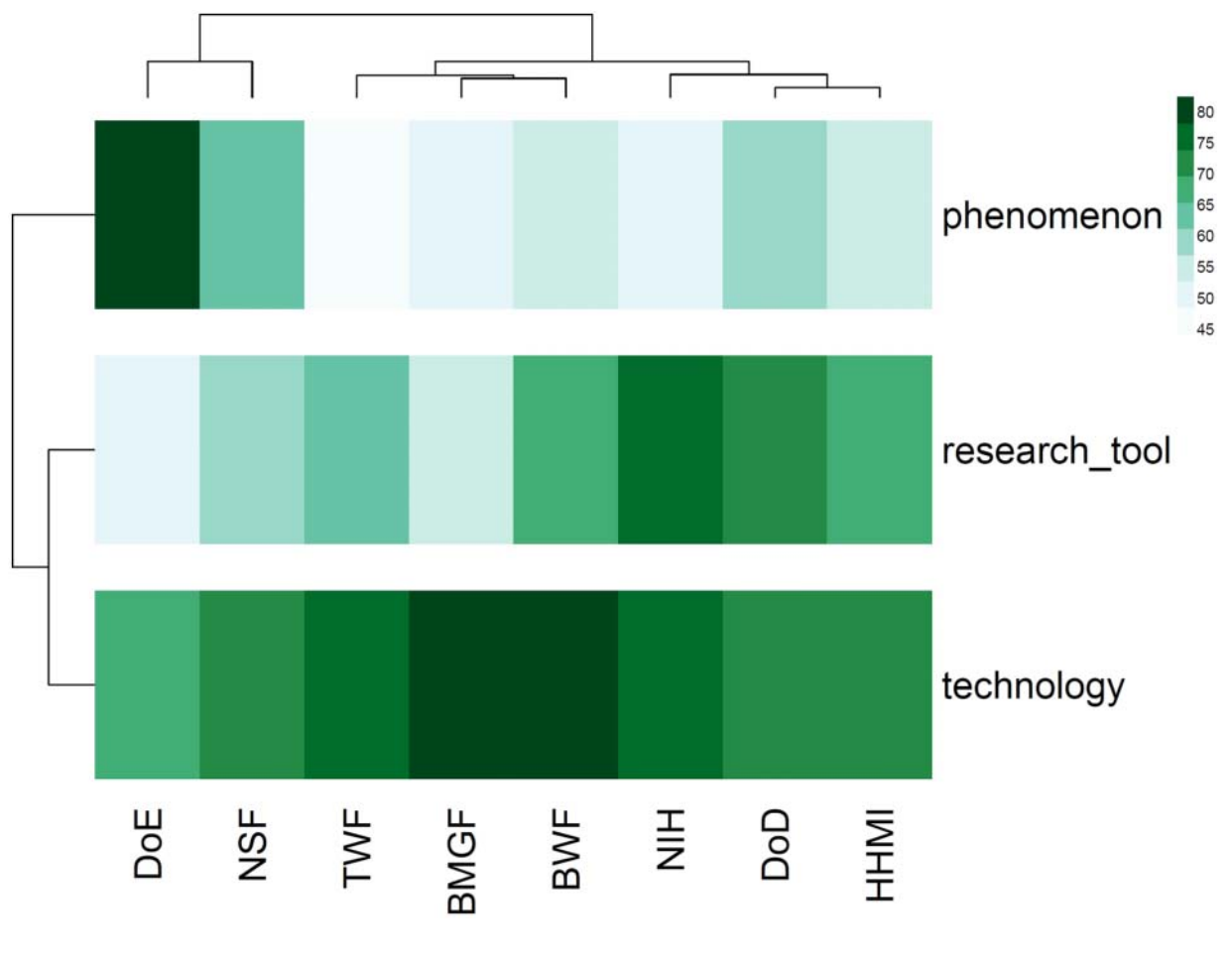

**Figure 2.** Heat map and clustering of coded categories versus philanthropic and government funders.

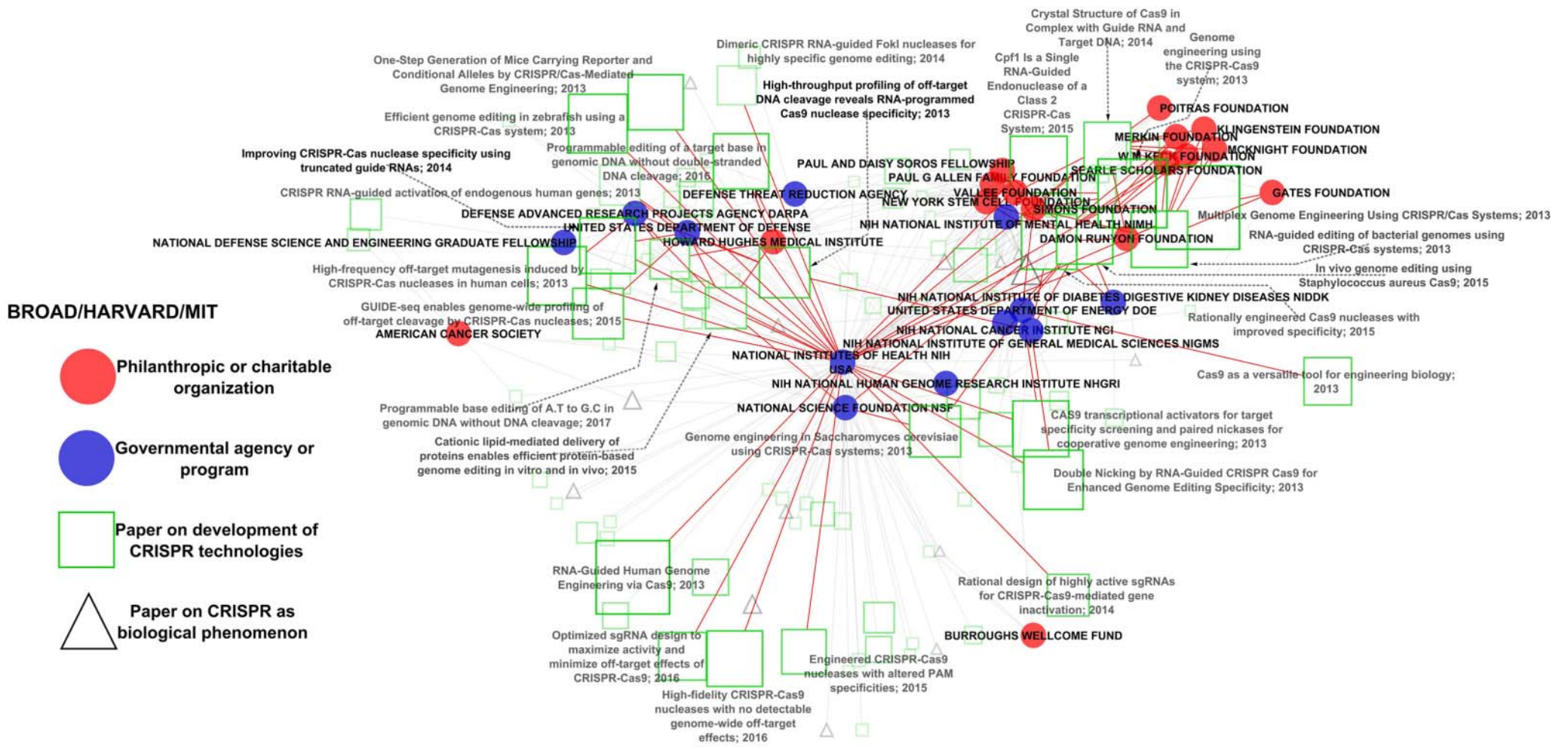

**Figure 3.** The cofunding network model of top cited CRISPR/Cas research in the Broad/Harvard/MIT system. This is a bimodal network model made of papers and funding sources. The size of the nodes representing papers is a function of the number of citations received. The edges point to which organizations funded which papers. Only the nodes representing the top most-cited papers in the network model (over 500 citations) are labeled with their title and year of publication and the red-lighted edges are linking them with their funding organization. Blue nodes represent governmental agencies while red nodes represent philanthropic/charitable organizations. It is important to specify that the aggregation of funding sources is limited by the information that the authors provide in the acknowledgment section of the papers. For example, there are authors who generally report funding from the National Institutes of Health, and other authors inform from which institute in particular they receive support.

In the case of the Broad/Harvard/MIT system, the investigation on the development of CRISPR/Cas technologies is grouped in three main clusters in the cofunding network model (Figure 3). The three clusters are supported by the NIH, which occupies a central position in the network model, but are cofunded by different sets or organizations. The cluster located in the lower right corner of the network model is formed by papers cofunded by the NIH and the Department of Energy (DoE); a second cluster (upper right corner) is cofunded by a set of philanthropic and governmental organizations; and the third cluster (upper left corner) is cofunded by the Howard Hughes Medical Institute (HHMI) together with a set of U.S. military organizations or programs (Figure 3). There is also an important set of papers exclusively funded by the NIH.

In the case of the University of California (UC), the cofunding network model (Figure 4) suggests that the top-cited investigations on CRISPR as a biological phenomenon, which is related to the discovery stage of investigation in which the basis of future technologies are built, were mostly supported by the U.S. Department of Energy and the National Science Foundation. A relatively smaller and less cited set of papers was cosupported by the Burroughs Wellcome Fund together with the National Institutes of Health (NIH; Figure 4). On the other hand, the investigations related to the development of CRISPR/Cas technologies at the UC were mostly supported by the NIH together with the National Science Foundation and the HHMI in two respective clusters of papers (Figure 4). There is also an important set of technological development papers funded by the NIH without the participation of other frequent governmental or philanthropic/charitable funding sources (Figure 4).

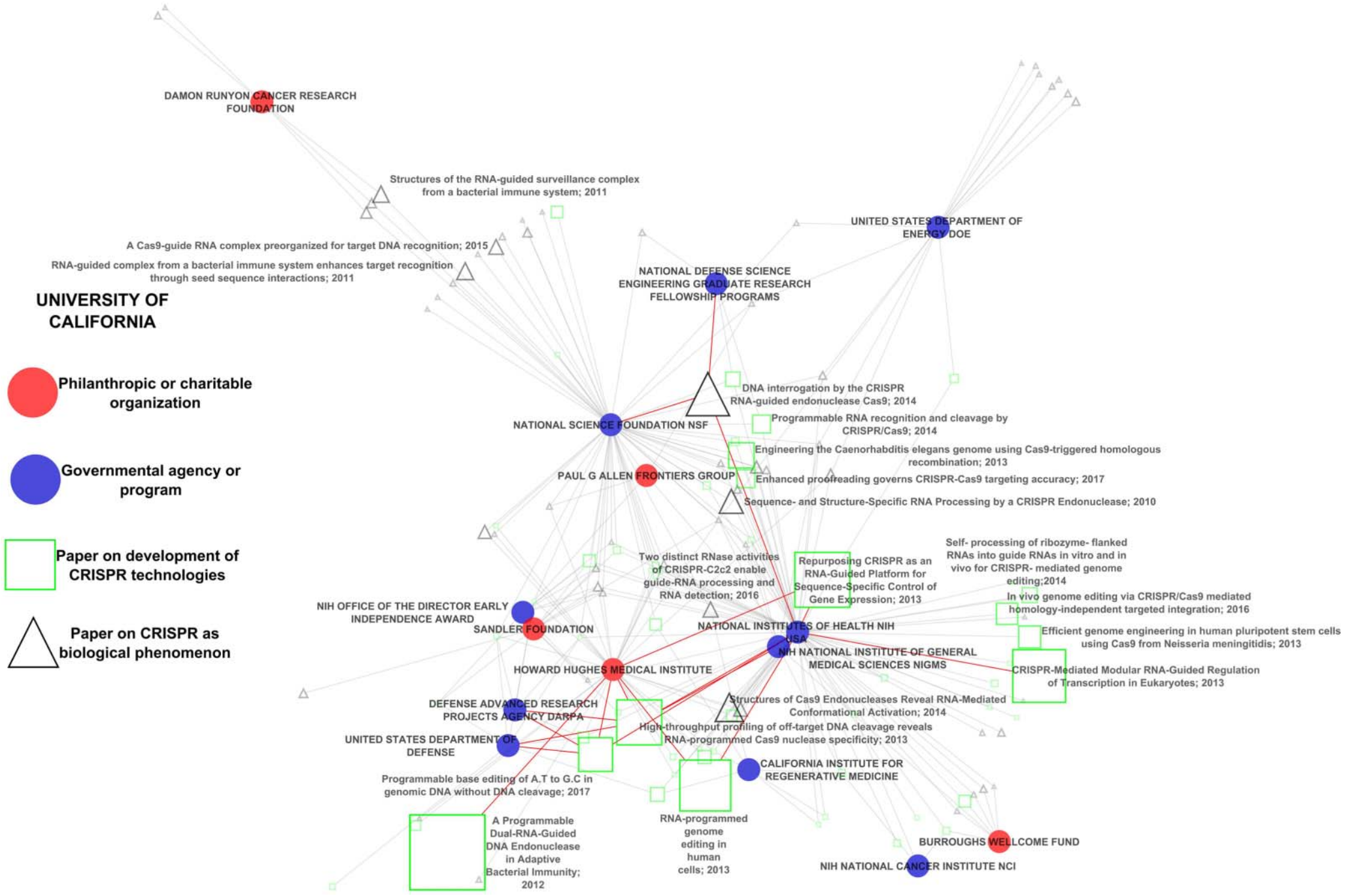

**Figure 4.** The cofunding network of top-cited CRISPR/Cas research in the UC system. This is a bimodal network model made of papers and funding sources. The size of the nodes representing papers is a function of the number of citations received. The edges point to which organizations funded which papers. Only the nodes representing the top most-cited papers in the network model (over 200 citations) are labeled with their title and year of publication and the red-lighted edges link them with their funding organizations. Blue nodes represent governmental agencies while red nodes represent philanthropic/charitable organizations. It is important to specify that the aggregation of funding sources is limited by the information that the authors provide in the acknowledgment section of the papers. For example, there are authors who generally report funding from the National Institutes of Health, and other authors inform from which institute in particular they receive support.

It is important to mention that while the papers authored by Jennifer Doudna are distributed throughout the bimodal network model, the papers of the three main authors associated with Broad/MIT/Harvard are concentrated in different regions of the bimodal network model. Feng Zhang's articles are funded primarily by the charitable/philanthropic cluster of organizations along with the NIH, while George Church's papers, on the network model, are cofunded primarily by NIH and DOE and David R. Liu's research is located in the cluster of papers cofinanced by HHMI in conjunction with the Department of Defense (DoD), the Defense Advanced Research Projects Agency (DARPA) and the Defense Threat Reduction Agency (DETRA).

## 4. DISCUSSION

The use of funding acknowledgment sections as a source of data for scientometrics studies has been extensively discussed in recent years (Kokol & Vošner, 2018; Paul-Hus, Desrochers, & Costas, 2016; Tang, Hu, & Liu, 2017). However, despite the enormous potential of using this source of data to investigate the impact of funding entities on the development of science and technology, some key considerations must be taken into account. First, our investigation was based on information provided by WoS, which began systematically collecting acknowledgments information in 2008 (Kokol & Vošner, 2018), becoming the de facto standard source for this type of investigation. In that sense, a comparative analysis of the three main bibliometric databases (WoS, PubMed, and Scopus) found that WoS outperforms the other databases in terms of the proportion of articles with funding information (Paul-Hus et al., 2016). Secondly, social sciences, humanities, and non-English language journals are underrepresented in both WoS and Scopus (Mongeon & Paul-Hus, 2016). In the case of WoS, specifically, the Arts & Humanities Citation Index (AHCI) content is not indexed for funding acknowledgment data and there are problems covering this information for non-English-language papers (Paul-Hus et al., 2016), although information on the financing of the papers indexed in AHCI has recently been added to WoS (Liu, Tang, & Hu, 2020). Third, a loss of funding information of 12% has been reported for WoS (Álvarez-Bornstein, Morillo, & Bordons, 2017). That is, 12% of the funding sources in the acknowledgment sections of papers are not captured in the WoS database. In the present investigation, even though WoS has systematized financing data, the name of each reported organization was carefully reviewed and errors were corrected. Previous studies reporting the use of funding acknowledgment sections focused on analyzing the impact of specific funding sources by using traditional metrics such as the number of citations per paper.

One way to understand the profiles of philanthropic and government funding sources for CRISPR research is to classify funding sources as targeted, top down, and nontargeted, bottom up. That is, there are, on the one hand, institutions that finance research following strategic criteria to reorient research efforts to satisfy specific knowledge and technology needs. An example of a targeted source of research funding would be the European Commission's FP7-health, which issues calls under priority research topics (Viergever & Hendriks, 2016). At the other extreme, there are organizations that offer their research funding through open competition based on academic merit. Such would be the case with HHMI (Tjian, 2015). Between these two extremes, there are organizations with mixed approaches to research funding, such as the UK Medical Research Council (Viergever & Hendriks, 2016). Our content analysis suggests that three of the top four sources of philanthropic funding for CRISPR research in the United States, the BWF, TWF, and BMGF, tend to specialize in the development of CRISPR as a technology. In the case of the BMGF and the BWF, their relative specialization in the development of CRISPR as a technology is to be expected, as both organizations have a science financing strategy aimed at solving the great challenges of society (Burroughs Wellcome Fund, 2021; Bill & Melinda Gates Foundation, 2022). TWF, for its part, focuses on the development of chemistry in the State of Texas (Welch Foundation, 2022). However, recently this foundation has shown interest in the subject of genome editing technologies (Doudna, 2019). In the case of the NSF, this federal agency of the United States has a legal mandate to support all fields of fundamental science and engineering (National Science Foundation, 2018). This would explain the high percentage of articles funded by the NSF that include terms related to both the study of CRISPR as a biological phenomenon and the development of CRISPR as a genomic technology. In the case of the Department of Energy, the high percentage of articles related to the study of CRISPR as a biological phenomenon can be explained in part by the research work that Jennifer Doudna carried out at the Lawrence Berkeley National Laboratory (Dabbar, 2021).

Our results show that the HIH, DoD, and HHMI maintain high percentages of articles related to the development of CRISPR technologies and their use as a biomedical research tool (Figure 2). The DoD is classified as a largely targeted research funding source (Viergever & Hendriks, 2016). The DoD's funding profile for CRISPR research can be explained in part by two strategic goals in relation to this technology. The first objective is related to making CRISPR a more secure technology by anticipating possible bioterrorist threats (Sanders, 2019). The second objective of the DoD in relation to the use of CRISPR is to protect the health of its personnel exposed to carcinogenic agents, and for this it requires investing in basic biomedical research (Glasgow, 2021). For its part, NIH's CRISPR research funding profile aligns well with its mission "to seek fundamental knowledge about the nature and behavior of living systems and the application of that knowledge to enhance health, lengthen life, and reduce illness and disability" (National Institutes of Health, 2017).

As far as we know, this is the first time that funding acknowledgments have been used to build and analyze the relationships between the structures of cofunding networks, the research level, and the type of funding sources. This methodology could allow policymakers, researchers, and granting agencies to identify whether different types of research funding (e.g., public, private, philanthropic, university-based) and different government agencies (e.g., NSF, NIH) tend to fund certain areas of scientific knowledge, how they influence one another, and what their impact is on highly influential science and technology development.

In the case of CRISPR/Cas, our results suggest that even though U.S. government agencies extensively supported all levels of CRISPR/Cas research in both institutional systems, philanthropic organizations have concentrated participation in cofunding the development of CRISPR/Cas technologies as opposed to basic biological phenomenon research. This is particularly true for the Broad/Harvard/MIT system, where research has clustered around particular research themes. Here, we observed relatively smaller philanthropic organizations clustering around similar developments, while larger organizations were more likely to fund projects independently of other charitable actors. This philanthropic funding comes on top of public research dollars, potentially assisting, though also altering, the trajectory of publicly funded science.

Using funding acknowledgments sections does not allow us to differentiate the impact of the different sources of funding on each research project. Further investigations are needed to analyze such differentiation. What our results do clearly show is that philanthropic organizations show a different behavior from government agencies when it comes to funding CRISPR research. Both the content analysis of the articles financed by the main sources of U.S. research funding and the analysis of funding networks at UC and the Broad/Harvard/MIT system show that unlike government agencies whose funding is more widespread, philanthropic organizations focus on research related to specific developments of CRISPR/Cas technologies. Particularly noteworthy is the concentration of numerous philanthropic organizations in funding research spearheaded by Feng Zhang to improve CRISPR/Cas technologies through the search for and development of new nucleases: enzymes which perform the critical "cutting" function of the CRISPR/Cas system (Figure 3).

Our results raise questions about the role of philanthropy in influencing predominantly publicly funded research trajectories and its potential contribution to the privatization of reward from that public investment. That is, while society as a whole finances the innovation process through government agencies—assuming most of the risks of investing in new areas of knowledge and technology—for-profit actors tend to participate in later stages (Laplane & Mazzucato, 2020; Mazzucato, 2015). Our results align with a model in which philanthropic organizations may play an important role in furthering the socialization of risk and the privatization of profits that comes from the basic/applied division of labor between the public and private sectors (McGoey, 2014). Critical studies on the governance of the innovation process in biotechnology, particularly the governance of CRISPR/Cas technologies, have largely overlooked the important role of U.S. charitable and philanthropic organizations as powerful actors that can redirect the trajectory of the development and application of genomic technologies in favor of specific interests or sectors of society. In that sense, it is fundamental to further the study of the interaction between transformative innovation and philanthropy. A first strategy would be to measure the impact of philanthropic grants by analyzing the expenditures of the sponsored projects. A consortium of 33 U.S. public research universities is moving in that direction by connecting the information on their sponsors, grants, project expenditures, and final products such as papers and patents (Owen-Smith, Lane et al., 2017). Unfortunately, neither the University of California system nor the Broad/Harvard/MIT system participate in this initiative. Another strategy would be to gather the views of researchers, administrators, and philanthropic foundations on the impact of such grants in the development of CRISPR/Cas technologies. Any strategy to deepen knowledge about the role of philanthropic foundations in the development of genomic editing technologies necessarily requires a commitment to transparency on the part of the various participating actors.

## ACKNOWLEDGMENTS

This research was supported by the Alexander von Humboldt Foundation. We thank our colleagues from Deutsche Zentrum für Hochschul-und Wissenschaftsforschung (Berlin, Germany) and El Colegio de la Frontera Sur (Chiapas, Mexico,) who provided insight and expertise that greatly assisted the research. This paper is our analysis and does not necessarily represent our colleagues' views.

## AUTHOR CONTRIBUTIONS

David Fajardo-Ortiz: Conceptualization, Data curation, Formal analysis, Funding acquisition, Investigation, Methodology, Project administration, Resources, Visualization, Writing—original draft, Writing—review & editing. Stefan Hornbostel: Conceptualization, Funding acquisition, Supervision. Maywa Montenegro de Wit: Conceptualization, Methodology, Writing—original draft. Annie Shattuck: Conceptualization, Methodology, Writing—original draft.

## COMPETING INTERESTS

The authors have no competing interests.

## FUNDING INFORMATION

This research was funded through an Alexander von Humboldt Stiftung (https://www.humboldt-foundation.de) postdoctoral grant to David Fajardo-Ortiz (grant number: 1195113). The funders had no role in study design, data collection and analysis, decision to publish, or preparation of the manuscript.

## DATA AVAILABILITY

The original source of information was WoS. The WoS data have been made available to DZHW and KU Leuven under a paid license. We are not allowed to redistribute WoS data used in this paper.